\documentclass{svjour3}                     
\smartqed  
\usepackage{graphicx}
 \usepackage{mathptmx}      
\usepackage{amssymb}
\usepackage{amsmath}
\usepackage{bm}
 \usepackage{lineno}
\usepackage{natbib}

\newcommand{\pd}{{\delta}}
\newcommand{\s}{{s}}

\newcommand{\spp}{p}
\newcommand{\pis}{{\tilde{\spp}}}

\newcommand{\bG}{\mathbf{G}}

\newcommand{\sS}{{\mathcal{S}}}

\newcommand{\bdelta}{{\bm{\delta}}}
\newcommand{\bs}{{\mathbf{s}}}
\newcommand{\ws}{{\tilde{w}}}

\newcommand{\nn}{\nonumber}

\newcommand{\be}{\begin{equation}}
\newcommand{\ee}{\end{equation}}

\newcommand{\bernoulli}{{\mathbb{B}}}
\newcommand{\tangent}{{\mathbb{T}}}
\newcommand{\euler}{{\mathbb{E}}}

\journalname{arXiv}
\begin{document}

\title{The link between segregation and phylogenetic diversity
}


\author{David Bryant         \and
        Steffen Klaere 
}


\institute{D. Bryant \at
	Allan Wilson Centre for Molecular Ecology and Evolution, {\em and}\\
             Dept. Mathematics and Statistics, University of Otago\\
             P.O. Box 56. Dunedin 9054, New Zealand.\\
              Tel.: +64-3-4797889, Fax: +64-3-4798427\\
              \email{david.bryant@otago.ac.nz}           
           \and
           S. Klaere \at
            Dept. Mathematics, University of Auckland\\
            Private Bag 9201, Auckland 1043, New Zealand.\\
            Tel.:+64-9-9235506, Fax:+64-9-3737457\\
            \email{s.klaere@math.auckland.ac.nz}
}

\date{Received: date / Accepted: date}

\maketitle

\begin{abstract}
We derive an invertible transform linking two widely used measures of species diversity:  phylogenetic diversity and the expected proportions of segregating (non-constant) sites. We assume a bi-allelic, symmetric, finite site model of substitution. Like the Hadamard transform of Hendy and Penny, the transform can be expressed completely independent of the underlying phylogeny. Our results bridge work on diversity from two quite distinct scientific communities.

\end{abstract}

\section{Introduction}
\label{sec:intro}

The quantification of biodiversity is one of the central conservation applications of  molecular ecology. There are, however, multiple ways to evaluate diversity, and different scientific communities have emphasised different measures. 

Among {\em population geneticists}, one of the two standard measures for diversity is the proportion of segregating sites ($\s$) \citep{watterson1975}. A site is segregating if it varies over the sampled taxa, and the proportion of segregating sites is often used when estimating population parameters. If $A$ is a subset of sampled taxa we let $s_A$ denote the probability that a site varies over the taxa in $A$. We note that the other standard population genetic measure of diversity, the expected pairwise divergence $\pi$,  is then just the average of $s_{A}$ over all subsets $A$ of size two.

Among {\em phylogeneticists} it is now standard to incorporate the phylogeny into assessments of diversity. Specifically, if $T$ is a phylogeny and $A$ is a subset of the taxa then the {\em phylogenetic diversity} $\delta_A$ of $A$ with respect to $T$ is the sum of the branch lengths in the smallest subtree of $T$ connecting $A$ \citep{faith1992}. One potential weakness of this measure is its dependence on a specific phylogeny, a problem that can be at least partially addressed by considering a distribution of trees \citep{minh2009,spillner2008,moulton2007}.

Here we consider the assessment of diversity using bi-allelic sites from the same haplotype block, that is, there is no recombination and all of the sites evolved on the same phylogeny $T$. We will show that in this instance the segregating site probabilities $s_A$ and the phylogenetic diversity measures $\delta_A$ are tightly linked. Specifically:
\begin{enumerate}
\item[(i)]
If we know the probabilities $s_A$ for all subsets $A$ of {\em even} cardinality then we can determine $s_A$ for subsets of any cardinality. Likewise, if we know $\pd_A$ for all subsets $A$ of {\em even} cardinality then we can determine $\pd_A$ for all subsets of any cardinality.
\item[(ii)] When the mutation rates are symmetric, then the segregating site probabilities and phylogenetic diversity probabilities are linked by an invertible transform
\be
\bdelta = -\bG \log (\bm{1} - \bG^{-1}\bs) \label{eq:transform}.
\ee
Here $\bG$ is a non-singular matrix independent of $T$, while $\bdelta$ and $\bs$ are vectors of phylogenetic diversity values and segregating site probabilities, both indexed by subsets of even cardinality. The entries $\bG_{AB}$ of $\bG$ are given by 
\begin{equation}
\bG_{AB} =  \begin{cases} 2^{-|A|+1}, &\mbox{ $B \subseteq A$;} \\ 0, & \mbox{ otherwise,}\end{cases}\end{equation}

\end{enumerate}
This transform can be used to relate phylogenetic diversity and segregating sites without any knowledge of the true phylogeny. 

The transform in equation~\eqref{eq:transform} is therefore a diversity analogue of the elegant Hadamard transform \citep{Hendy89,Hendy89a,hendy1993,hendy2005} which   maps branch weights in a tree to pattern probabilities. As could be expected, the two transforms are closely related mathematically. We make use of Hendy's path-set approach when we prove the correctness of our transform. Note that a version of \eqref{eq:transform} can be obtained by transforming diversities to split weights (using Theorem~\ref{thm:div2split} below), applying the Hadamard transform, then transforming pattern probabilities to segregating site probabilities. This approach introduces an undesirable asymmetry in the indexing of the transform because of the need to specify a reference taxon in the Hadamard transform. We were unable to remove this asymmetry, and instead derived a more direct result.


\begin{figure}[htbp] 
   \centering
   \includegraphics[width=2in]{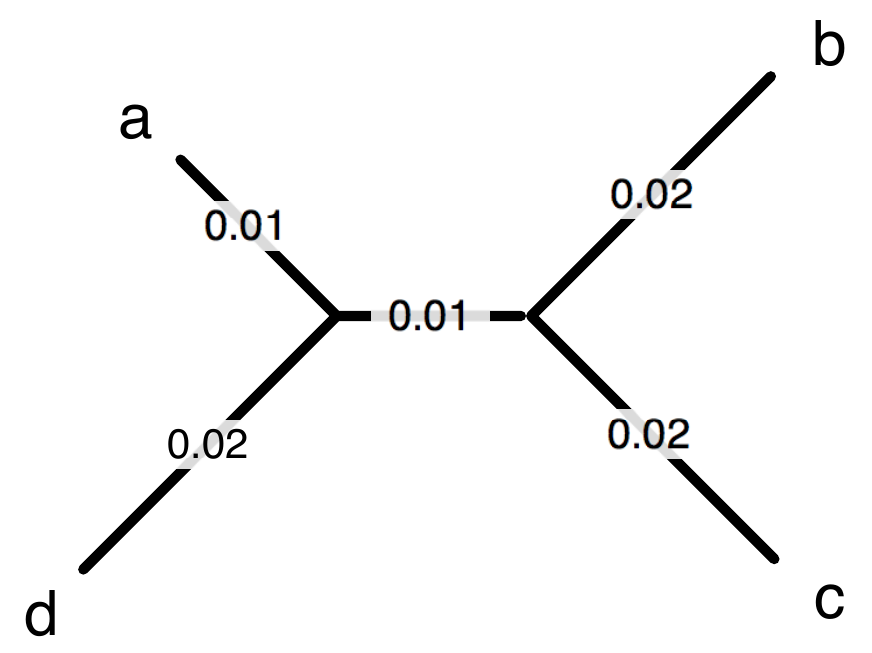} 
   \caption{Four taxon tree with branch lengths indicated.}
   \label{fig:fourTaxon}
\end{figure}

To illustrate the result, consider the four taxon phylogeny in Figure~\ref{fig:fourTaxon}. A pattern is simulated by selecting an (arbitrary) root node, choosing the state at the root uniformly at random, and then evolving the states along branches away from the root. For each branch, if $t$ is the length of the branch then the probability of a state change along that branch is $\frac{1}{2}(1 - e^{-2t})$.  Under this symmetric model, the probabilities of the various site patterns are
\begin{align*}
p_{0000} &= p_{1111}  = 0.4619,\\
 p_{0001} &= p_{1110}  =   0.0047,\\
 p_{0010} &= p_{1101}  =   0.0093,\\
p_{0011} &= p_{1100}  =    0.0003,\\
p_{0100} &= p_{1011}  =    0.0093,\\
p_{0101} &= p_{1010}  =    0.0003,\\
 p_{0110} &= p_{1001}  =   0.0049,\\
p_{0111} &= p_{1000}  =    0.0093.
\end{align*}
The phylogenetic diversity values for any subset can be computed directly from the tree by summing the appropriate branch lengths, while the segregating site probabilities can be computed from the site pattern  probabilities. In this way we obtain
\begin{eqnarray*}
\bdelta & = & (\delta_{ab},\delta_{ac},\delta_{ad},\delta_{bc},\delta_{bd},\delta_{cd},\delta_{abcd})' \\
& = & (0.04, 0.04, 0.03, 0.04, 0.05, 0.05, 0.08)';\\
\bs & = &  (s_{ab},s_{ac},s_{ad},s_{bc},s_{bd},s_{cd},s_{abcd})' \\
& = & (0.0384,    0.0384,    0.0291,    0.0384,    0.0476,    0.0476,    0.0762)'.
\end{eqnarray*}
Our main result says that these two vectors are related by \eqref{eq:transform}, where in this case 
\[G = \begin{pmatrix}
0.5&0&0&0&0&0&0\\
0  &  0.5&0&0&0&0&0\\
0&0  &  0.5&0&0&0&0\\
0&0&0 &   0.5&0&0&0\\
0&0&0&0 &   0.5&0&0\\
0&0&0&0&0 &   0.5&0\\
    0.125 &   0.125 &   0.125 &   0.125 &   0.125  &  0.125 &   0.125
    \end{pmatrix}.
    \]
   A quick calculation validates the formula in this instance. 
%


The paper is structured as follows. The next section introduces some basic properties relating segregating site probabilities to site pattern probabilities. Section 3 shows that these properties have analogues relating phylogenetic diversity and branch lengths. Section 4 provides the bridge connecting segregating sites and phylogenetic diversity.


\section{Properties of segregating site probabilities}
\label{sec:basis}

Let $X$ denote the set of taxa. For any $A \subseteq X$, let $\spp_A$ denote the probability that, for a given site, the taxa in $A$ have state $1$ and the taxa in $\overline{A} = X \setminus A$ have state $0$. At this point, we will not be making any assumptions about the probabilities $\spp_A$ beyond that they are non-negative and $\sum_{A \subseteq X} \spp_A = 1$. 

Let $\pis$ denote the symmetrised version of $\spp$, defined by
\[
\pis_A = (\spp_A + \spp_{\overline{A}})/2.
\]
A site is segregating over a subset $A \subseteq X$ if it is not constant over $A$. Hence if $B$ is the set of taxa with state $1$ for some site then the site is segregating if and only if $A \cap B$ and $A \cap \overline{B}$ are both nonempty. We therefore define the collection
\[
\sS_A = \left \{B : A \cap B \neq \emptyset \mbox{ and } A \cap \overline{B} \neq \emptyset \right\}
\]
so that the probability that a randomly chosen site is segregating over $A$ is given by
\be \label{eq:probseg}
\s_A = \sum_{B \in \sS_A} \spp_B.
\ee
Since $B \in \sS_A$ if and only if $\overline{B} \in \sS_A$ we also have
\be \label{eq:probsegs}
\s_A = \sum_{B \in \sS_A} (\spp_B + \spp_{\overline{B}})/2 = \sum_{B \in \sS_A} \pis(B).
\ee
There are $2^{n-1} -1$ degrees of freedom for the symmetrised probabilities $\pis$. In contrast, there are $2^{n} - n - 1$ values for $s$, one for every subset with cardinality at least two. Hence there must therefore be a great deal of redundancy in the segregating site probabilities. We show here that the probabilities $\s_A$ for all $A \subseteq X$ are determined by the probabilities $\s_A$ for $A$ with even cardinality, so that segregating site probabilities have the same degrees of freedom as the symmetrised site pattern probabilities $\pis$.

We will make use of the tangent numbers $\tangent_k$, which are defined by the power series
\[
\tanh(x) = \sum_{k=0}^\infty \tangent_k \frac{x^k}{k!}
\]
and are related to the better known Bernoulli numbers via the identity
\[
\tangent_k=2^{k+1}(2^{k+1}-1)\bernoulli_{k+1}/(k+1),
\]
see \cite[pg. 6--7]{cohen2007}.

\begin{theorem}\label{thm:seg.even}
\begin{enumerate}
\item For all $A \subseteq X$,
\begin{equation}
\s_A = \sum_{B \subseteq A} (-1)^{|B|} \s_B.
\end{equation}
\item If $|A|$ is odd then $s_A$ is determined by $s_B$ values for $|B|$ even, by
\begin{equation}s_A = \sum_{\substack{ B \subset A\\ |B| even}} \frac{\tangent_{|A| - |B|}}{2^{|A|-|B|}} \s_B. \label{eq:sRecurse}
\end{equation}
\end{enumerate}
\end{theorem}

\begin{proof}
1. Suppose that $A \subseteq X$. Then we get
\begin{eqnarray*}
\sum_{B \subseteq A} (-1)^{|B|} \s_B & = & \sum_{\substack{B \subseteq A\\|B| even}} \s_B - \sum_{\substack{B \subseteq A\\|B| odd}} \s_B \\
& = &  \sum_{\substack{B \subseteq A\\|B| even}} \sum_{C \in \sS_B} \pis(C) -  \sum_{\substack{B \subseteq A\\|B| odd}} \sum_{C \in \sS_B} \pis(C).
\end{eqnarray*}
Consider any $C \subseteq X$. If $A \cap C = \emptyset$ or $A \cap \overline{C}=\emptyset$ then there is no $B \subseteq A$ such that $C \in \sS_B$. 
Otherwise, suppose that $C \in \sS_A$, let $k = |A \cap C|$ and $l = |A \setminus C|$. 

The number of subsets $B \subseteq A$ such that $|B|$ even and $C \in \sS_B$ is then 
\[
2^{k-1} 2^{l-1} + (2^{k-1}-1)(2^{l-1}-1) = 2^{k+l-1} - 2^{k-1} - 2^{l-1} + 1
\]
which follows by looking at the possible (non-empty) intersections of $B$ with $A \setminus C$ and $A \cap C$.  The number of subsets $B \subseteq A$ such that $|B|$ even and $C \in \sS_B$ is
\[
2^{k-1}(2^{l-1}-1) + (2^{k-1}-1)2^{l-1} = 2^{k+l-1} - 2^{k-1} - 2^{l-1}.
\]
Hence
\begin{eqnarray*}
\sum_{B \subseteq A}(-1)^{|B|}\s_B&=&\sum_{C \in \sS_A}(2^{k+l-1}-2^{k-1}-2^{l-1}+1)\pis(C)-\sum_{C\in\sS_A}(2^{k+l-1}-2^{k-1}-2^{l-1})\pis(C)\\
&=&\sum_{C\in\sS_A}\pis(C)\\
&=&\s_A.
\end{eqnarray*}
2. The result holds trivially when $|A| = 1$. Suppose $|A|=2r+1$ and the result holds for sets with cardinality less than $|A|$. Then from part 1,
\begin{eqnarray*}
\s_A &=&   \frac{1}{2} \sum_{\substack{B \subset A\\|B| even}}\s_B - \frac{1}{2}\sum_{\substack{C \subsetneq A\\|C| odd}}\s_C \\
& = &  \frac{1}{2} \sum_{\substack{B \subset A\\|B| even}}\s_B - \frac{1}{2}  \sum_{\substack{C \subset A\\|C| odd}}\sum_{\substack{ B \subset C\\ |B| even}} \frac{\tangent_{|C| - |B|}}{2^{|C|-|B|}} \s_B \\
& = &  \frac{1}{2} \sum_{k=0}^r \sum_{\substack{B \subseteq A\\|B| = 2k}} \left(1 -  \sum_{j=k}^{r-1}   \sum_{\substack{C: B\subset C \subset A\\|C| =2j+1}}  \frac{\tangent_{2(j-k)+1}}{2^{2(j-k)+1}}   \right)\s_B\\
& = &  \frac{1}{2} \sum_{k=0}^r \sum_{\substack{B \subseteq A\\|B| = 2k}} \left(1 -  \sum_{j=0}^{r-k-1} \binom{2(r-k)+1}{2j+1} \frac{\tangent_{2j+1}}{2^{2j+1}}   \right)\s_B
\end{eqnarray*}
The identity
\[
\frac{\tangent_{2(r-k)+1}}{2^{2(r-k)+1} }= \frac{1}{2} - \frac{1}{2} \sum_{j=0}^{r-k-1}  \binom{2(r-k)+1}{2j+1} \frac{\tangent_{2j+1}}{2^{2j+1}}  
\]
can be proven by substituting the power series
\[
\tanh(t)=\sum_{k=0}^\infty\frac{t^{2k+1}}{(2k+1)!}\tangent_{2k+1}
\]
into the expression
\[
\tanh(t/2)=\frac{\text{e}^t-1}{\text{e}^t+1}
\]
and equating coefficients.
\end{proof}

\vspace{5mm}
Having shown that the $s$ values and $\pis$ values have the same degrees of freedom, we now give an explicit formula for the $\pis$ values in terms of all $s_B$ values. A consequence of this result is that, in the case of symmetric mutation rates, the segregating site probabilities are (trivially) sufficient statistics for inferring trees of population parameters.

\begin{theorem}\label{thm:inverse}
For all $B \subseteq X$ we have
\be
\pis_B = \frac{1}{2} \sum_{A:B \subseteq A} (-1)^{|A| - |B| + 1} \s_A.
\ee
\end{theorem}

\begin{proof}
Suppose that $\emptyset \subsetneq B \subsetneq X$. Every site segregating over $X \setminus B$ is segregating over $X$. Conversely, a site is segregating over $X$ but not over $X \setminus B$ exactly if all the taxa with ones for the site are contained in $B$ or all the taxa with ones are contained in $X \setminus B$. Hence
\[
\s_X - \s_{X \setminus B} = \sum_{V \subseteq B} 2 \pis_{X \setminus V}.
\]
Applying M\"obius inversion we have for all $\emptyset\ne V\subsetneq X$ that
\begin{eqnarray*}
\pis_{V} & = & \frac{1}{2} \sum_{B \subseteq V} (-1)^{|V| - |B|} \left(\s_X - \s_{X \setminus B} \right) \\
& = & \frac{1}{2}\sum_{B \subseteq V} (-1)^{|V| - |B|+1} \s_{X \setminus B}\\
& = & \frac{1}{2}\sum_{A \supseteq U} (-1)^{|A|-|U|+1} \s_A,
\end{eqnarray*}
this last step given by the substitution $A = X \setminus B$.
\end{proof}


\section{Properties of phylogenetic diversities}

Let $T$ be a phylogeny with branch lengths and leaf set $X$. We defined the phylogenetic diversity of a subset $A \subseteq X$ as the length of the smallest subtree of $T$ connecting the leaves in $A$, where the length of a subtree is the sum of all its branch lengths. \cite{minh2006} and  \cite{moulton2007} showed that phylogenetic diversity could also, conveniently, be defined in terms of splits.

A {\em split} of $X$ is a bipartition of $X$ into two parts, where here we will also permit the trivial bipartition $\emptyset | X$. We regard $U|V$ and $V|U$ as the same split and let $\Sigma_X$ denote the set of all splits of $X$.
Deleting an edge $e$ in a phylogenetic tree induces a split $\sigma_e$ of $X$ given by the leaf sets of the resulting two components. The set of all splits of a tree $T$ obtained in this way is denoted $\Sigma(T)$. A {\em split weight function} is a map $w:\Sigma_X \rightarrow \Re_{\geq 0}$. The split weight function for a phylogenetic tree with branch lengths is defined by
\be
w_{U|V} = \begin{cases} b_e & \mbox{ if $U|V = \sigma_e$ for some edge $e \in E(T)$ with length $b_e$} \\
0& \mbox{ otherwise.} \end{cases} 
\ee
See Chapter 3 of \cite{semple2003} for more on splits and their uses in phylogenetic combinatorics.

\cite{moulton2007} and \cite{minh2009} showed that if $w$ is the split weight corresponding to a tree with taxon set $X$, and $A \subseteq X$, then 
\begin{equation}
\pd_A = \sum_{\substack{U|V \in \Sigma_X\\U \cap A \neq \emptyset\\V \cap A \neq \emptyset}} w_{U|V}. \label{eq:splitdiv} 
\end{equation}
Actually, this formulation applies to any split weight function, allowing an extension of phylogenetic diversity to phylogenetic networks.  

As we shall see, the mathematics of phylogenetic diversity is in many ways dual to the mathematics of segregating sites. As before, we define for $A \subseteq X$ the set $\sS_A = \{B \subseteq X: B \cap A \neq \emptyset, \overline{B} \cap A \neq \emptyset\}$. Let $\ws$ be the function from subsets of $X$ to real numbers given by
\[
\ws_A = w_{A | \overline{A}}/2.
\]
We then have from \eqref{eq:splitdiv} that
\be 
\pd_A = \sum_{B \in \sS_A} \ws_B. \label{eq:splitdivs}
\ee
This is, of course, the same as \eqref{eq:probsegs} with a change of labels. Hence we immediately obtain the analogues to Theorem~\ref{thm:seg.even} and Theorem~\ref{thm:inverse}.
 
\begin{theorem}
\begin{enumerate}
\item For all $A \subseteq X$,
\be
\pd_A = \sum_{B \subseteq A} (-1)^{|B|} \pd_B.
\ee
\item
For all $A \subseteq X$ with odd cardinality,
\be
\pd_A = \sum_{B \subset A: |B| \mbox{ even}} \frac{\tangent_{|A|-|B|}}{2^{|A|-|B|}} \pd_B.
\ee
\end{enumerate}
\end{theorem}
 
\begin{theorem}\label{thm:div2split}
For all $B \subseteq X$ we have
\be
\ws_B = \frac{1}{2} \sum_{A:B\subseteq A}(-1)^{|A|-|B|+1}\pd_A.
\ee
\end{theorem}

\section{An invertible transform for segregating sites and diversities}

We have seen how the phylogenetic diversity and segregating site probabilities have a similar structure. Here we formally establish a link between the two, one that works irrespective of the underlying phylogeny. The transform takes a vector of $s_A$ values and returns a vector of $\pd_A$ values, and does so via intermediate values $\gamma$ and $\mu$ which we now define.

\begin{theorem} \label{thm:introgammamu}
\begin{enumerate}
\item For each $A \subseteq X$ define 
\[
\gamma_A := \begin{cases}  
\sum\limits_{B:|A \cap B| odd} \pis_B & \mbox{ if $|A|$ is even;} \\ 
0 & \mbox{ if $|A|$ is odd.} 
\end{cases}
\]
Then
\begin{eqnarray}
\gamma_A & = & \sum_{B \subseteq A} (-2)^{|B|-2} s_B; \label{eq:gamma2s}\\
s_B & = & \frac{1}{2^{|A|-2}} \sum_{A \subseteq B} \gamma_A. \label{eq:s2gamma}
\end{eqnarray}
\item For each $A \subseteq X$ define
\[
\mu_A := \begin{cases}  
\sum\limits_{B:|A \cap B| odd} \ws_B & \mbox{ if $|A|$ is even;} \\ 
0 & \mbox{ if $|A|$ is odd.} 
\end{cases}
\]
Then
\begin{eqnarray}
\mu_A & = & \sum_{B \subseteq A} (-2)^{|B|-2} \delta_B; \label{eq:delta2mu}\\
\delta_B & = & \frac{1}{2^{|B|-2} }\sum_{A \subseteq B} \mu_A. \label{eq:mu2delta}
\end{eqnarray}
\end{enumerate}
\end{theorem}

\begin{proof}
Substituting \eqref{eq:probsegs} into the right hand side of \eqref{eq:gamma2s} we obtain
\begin{eqnarray*}
\sum_{B \subseteq A} (-2)^{|B|-2} s_B & = & \sum_{B \subseteq A} (-2)^{|B|-2} \sum_{C \in \sS_B} \pis_C\\
 & = & \sum_{C \subseteq X} \left(    \sum_{\substack{B \subseteq A:\\ B \cap C, \,\,B \cap \overline{C} \neq \emptyset}} (-2)^{|B|-2}  \right) \pis_C .
\end{eqnarray*}
The number of subsets $B \subseteq A$ of cardinality $|B| = k$ such that $B \cap (A \cap C) \neq \emptyset$ and $B \cap (A \setminus C) \neq \emptyset$ equals $\binom{|A|}{k} - \binom{|A \cap C|}{k} - \binom{|A \setminus C|}{k}$.  By applying the binomial theorem three times, we obtain 
\begin{eqnarray}
\sum_{\substack{B \subseteq A:\\ B \cap C, \,\,B \cap \overline{C} \neq \emptyset}} (-2)^{|B|-2} & = & \frac{1}{4}\sum_{k=2}^{|A|} (-2)^k \left(\binom{|A|}{k} - \binom{|A \cap C|}{k} - \binom{|A \setminus C|}{k} \right) \nn \\
 & = & \frac{1}{4}\big(1+(-1)^{|A|}-(-1)^{|A\cap C|}-(-1)^{|A\setminus C|}\big). \label{eq:cancels}
\end{eqnarray}
If $|A|$ is odd, or if $|A|$ is even and $|A \cap C|$ is even, then \eqref{eq:cancels} becomes zero. If $|A|$ is even and $|A \cap C|$ is odd, then \eqref{eq:cancels} evaluates to $1$. This proves \eqref{eq:gamma2s}.  The inverse relation \eqref{eq:s2gamma} now follows by applying M\"{o}bius inversion to \eqref{eq:gamma2s}. 

The identities \eqref{eq:delta2mu} and \eqref{eq:mu2delta} are proved in the same way.
\end{proof}

\vspace{5mm}Combining Theorem~\ref{thm:introgammamu} and Theorem~\ref{thm:seg.even} we see that for each even cardinality set $A$, the value $\gamma_A$ is a linear function of the values $s_B$ with $|B|$ even. Likewise, each value $\mu_A$ is a linear function of the values $\delta_B$ with $|B|$ even. The following theorem makes the relationship explicit. We make use of the Euler numbers $\euler_k$, which are defined by the generating function
\[
\frac{1}{\cosh(x)} = \frac{2}{e^x+e^{-x}} = \sum_{k=0]}^\infty \euler_k \frac{x^k}{k!},
\]
see \cite[pg. 7]{cohen2007}.

\begin{theorem} \label{thm:gammaEven}
If $|A|$ is even then 
\begin{eqnarray}
\gamma_A & = & \frac{1}{4} \sum_{\substack{B \subseteq A\\|B| even}} 2^{|B|} \euler_{|A| - |B|} s_B,\\
\mu_A & = &  \frac{1}{4} \sum_{\substack{B \subseteq A\\|B| even}} 2^{|B|} \euler_{|A| - |B|} \delta_B.
\end{eqnarray}
\end{theorem}

\begin{proof}
Applying Theorem~\ref{thm:introgammamu} and Theorem~\ref{thm:seg.even} we obtain
\begin{eqnarray*}
\gamma_A & = & \sum_{B \subseteq A} (-2)^{|B|-2} s_B\\
&=&\sum_{\substack{B \subseteq A\\|B|even}} 2^{|B|-2} s_B - \sum_{\substack{C \subseteq A\\|C| odd}} 2^{|C|-2} s_C \\
& = &  \sum_{\substack{B \subseteq A\\|B|even}} 2^{|B|-2} s_B - \sum_{\substack{C \subseteq A\\|C| odd}} \sum_{\substack{B \subseteq C \\ |B| even}} 2^{|C|-2} \frac{\tangent_{|C|-|B|}}{2^{|C|-|B|} }s_B \\
& = &  \sum_{\substack{B \subseteq A\\|B|even}} \left[ 1 - \sum_{\substack{C:B \subseteq C \subseteq A \\ |C| odd} }\tangent_{|C|-|B|} \right]2^{|B|-2}s_B.
\end{eqnarray*}
If $|A| = 2r$ and $|B| = 2k$ then 
\begin{equation} 
1 - \sum_{\substack{C:B \subseteq C \subseteq A \\ |C| odd}} \tangent_{|C|-|B|}   =  1 - \sum_{j=k}^{r-1} \binom{2(r-k)}{2(j-k)+1}  \tangent_{2(j-k)+1}.
\label{eq:almostEuler}
\end{equation}
 Here we can apply the identity
\begin{equation}
\sum_{j=0}^n \binom{n}{k} \tangent_k = -\euler_{n},\label{eq:EulerIden}
\end{equation}
which is proven by substituting the generating functions
\begin{eqnarray*}
\sum_{k=1}^\infty \tangent_k \frac{x^k}{k!} & = & \tanh(x) \\
\sum_{k=0}^\infty \euler_k \frac{x^k}{k!} & = & \frac{2}{e^x + e^{-x}} 
\end{eqnarray*} 
into the expression
\[
(1-\tanh(x))e^x = \frac{2}{e^x + e^{-x}}.
\]
Applying \eqref{eq:EulerIden} to \eqref{eq:almostEuler}, and using the fact that $\tangent_0=-1$ and $\tangent_{2k} = 0$ for $k \geq 1,$ we obtain
\[
\sum_{j=k}^{r-1} \binom{2(r-k)}{2j+1-2k}\tangent_{2(j-k)+1}  = 1-\euler_{2(r-k)},
\]
proving the theorem.
\end{proof}

The final link in the transform is the map between $\gamma$ and $\mu$. Let $T$ be the underlying phylogenetic tree, so that $\delta_A$ is the length of the minimal subtree connecting $A$ in $T$, and $s_A$ is the probability that a site generated on $T$ is not constant over $A$. 

A {\em path-set} of $T$ is the set of edges in the disjoint union of a set of leaf-to-leaf paths of $T$.  Path-sets were introduced by \cite{hendy1993} when proving the correctness of the Hadamard transform, and we will make use of several of their results  \citep[see also][]{hendy2005}.

Consider a path-set with set of endpoints $A$, noting that a path-set is uniquely determined by its set of endpoints. The sum of edge lengths in the path-set, or the {\em length} of the path-set equals the sum of the split weights for all splits $U|V$ in the tree such that $|U \cap A|$ and $|V \cap A|$ are both odd. Hence, the length of the path-set is $\mu_A$. 

The probability that an odd number of taxa in $A$ have state $1$ is the sum of probabilities $\pis_U$ over all $U$ with $|U \cap A|$ odd. Hence, this probability equals $\gamma_A$. 

The core of the Hadamard transform is a formula connecting the length of a  path-set (in our case, $\mu_A$) and the probability that a site assigns $1$ to an odd number of taxa in the endpoints of the path-set (in our case, $\gamma_A$). From \cite{hendy1993} we obtain
\begin{equation}
\mu_A = -\frac{1}{2} \log(1 - 2 \gamma_A). \label{eq:pathset} 
\end{equation}
We now have invertible transforms from $\delta$ to $\mu$ to $\gamma$ to $s$. The following theorem makes the composite transform explicit.

\begin{theorem}
Let $\bdelta$ and $\bs$ denote the vector of $\delta_A$ and $s_A$ values, where $A$ ranges over non-empty subsets of $X$ with even cardinality. Then
\[\bdelta = -\bG \log (\bm{1} - \bG^{-1}\bs),\]
where
\[\bG_{AB} =  \begin{cases} 2^{-|A|+1}, &\mbox{ $B \subseteq A$;} \\ 0, & \mbox{ otherwise,}\end{cases}\]
and
\[(\bG^{-1})_{AB} =  \begin{cases} 2^{|B|-1} \euler_{|A|-|B|}, &\mbox{ $B \subseteq A$;} \\ 0, & \mbox{ otherwise.}\end{cases}\]
\end{theorem}
\begin{proof}
From Theorem~\ref{thm:gammaEven}, the above review of path-set results in \cite{hendy1993},  and Theorem~\ref{thm:introgammamu} , we have for all $A \subseteq X$ with $|A|$ even,
\begin{eqnarray*}
\gamma_A & = & \frac{1}{4} \sum_{\substack{B \subseteq A\\|B| even}} 2^{|B|} \euler_{|A| - |B|} s_B \\
\mu_A & = & -\frac{1}{2} \log(1 - 2 \gamma_A) \\
\delta_A & = & \frac{1}{2^{|A|-2} }\sum_{\substack{B \subseteq A\\|B| even}} \mu_B.
\end{eqnarray*}
\end{proof}

\begin{acknowledgements}
This research was supported by a Marsden grant, the University of Auckland faculty of Science grant, and the Allan Wilson Centre for Molecular Ecology and Evolution. 
\end{acknowledgements}

\bibliographystyle{spbasic}      
\bibliography{seg_div}   


\end{document}